\documentclass[runningheads]{llncs}
\usepackage{graphicx}
\usepackage{hyperref}
\usepackage{amsmath, amssymb} 
\newtheorem{assumption}{Assumption}
\usepackage[most]{tcolorbox}
\begin{document}
\title{Decentralized Governance of Stablecoins with Closed Form Valuation}
\titlerunning{Decentralized Governance of Stablecoins with with Closed Form Valuation}
%
\author{Lucy Huo\textsuperscript{1} \and Ariah Klages-Mundt\textsuperscript{1} \and Andreea Minca\textsuperscript{1} \and Frederik Christian M\"unter\textsuperscript{2} \and Mads Rude Wind\textsuperscript{2}}
\institute{Cornell University \and Copenhagen University}
\authorrunning{Huo et al.}
%
%
%

\maketitle

\begin{abstract}
	We model incentive security in non-custodial stablecoins and derive conditions for participation in a stablecoin system across risk absorbers (vaults/CDPs) and holders of governance tokens. We apply option pricing theory to derive closed form solutions to the stakeholders' problems, and to value their positions within the capital structure of the stablecoin. We derive the optimal interest rate that is incentive compatible, as well as conditions for the existence of equilibria without governance attacks, and discuss implications for designing secure protocols.
\end{abstract}


\section{Introduction}

Decentralized finance (DeFi) protocols are often described as either utopian systems of aligned incentives or dystopian systems that incentivize hacks and exploits.
These incentives, however, are thus far sparsely studied formally, especially around the governance of DeFi applications, which determine how they evolve over time.
Unlike in traditional companies, governance in DeFi is meant to be transparent and openly auditable through smart contracts on a blockchain.
The aim is often to incentivize good governance without relying on legal recourse, setting it apart from corporate finance.
While some DeFi applications are immutable, with change impossible, most have some flexibility to parameters (such as fees and price feeds, or ``oracles''), and often the entire functionality can be upgraded.
Control is often placed in the hands of a cooperative of governance token holders who govern the system.
This cooperative, however, is known to face perverse incentives, both theoretically (e.g., \cite{klagesmundt2019vuln,zoltu2019,gudgeon2020decentralized}) and often in practice (e.g., \cite{hack:compounder,rekt2021paid}) to steal (or otherwise extract) value.

{\it Related work.}
These incentives to deviate from the best interest of the protocol and its users are termed \emph{governance extractable value} (GEV) \cite{lee2021gov,werner2021sok}.
While there is a body of work on blockchain governance (e.g.~\cite{reijers2016governance,beck2018governance,lee2020political}), DeFi governance is sparsely studied.
\cite{klages2020stablecoins} proposed a framework for modeling DeFi governance that extends capital structure models from corporate finance (see \cite{dybvig1991capital,myers1984corporate}).
Equilibria in these models are not yet formally studied.
In this paper, we incorporate closed form valuation into the framework proposed in \cite{klages2020stablecoins} and characterize the equilibria around interest rate policies (and how closely these lead to stability) and governance attacks in non-custodial stablecoins.

{\it Non-custodial stablecoins.}
Non-custodial stablecoins are implemented as smart contracts using on-chain collateral, which are not controlled by a responsible party \cite{bullmann2019search}.
We briefly introduce the core components of these stablecoins and refer to \cite{bullmann2019search,klages2020stablecoins} for further details.

We focus on exogenous collateral, whose price is independent of the stablecoin system and which has proven  able to maintain the peg in the long run, see e.g. \cite{bullmann2019search}.
Stablecoin issuance is initiated by a user creating a collateralized debt position (CDP) using a ``vault''. The user transfers collateral, e.g., ETH, to the vault, which can mint an amount of stablecoins up to the minimum collateralization level. 
This leveraged position can be used in multiple ways, e.g. to spend the stablecoin or invest in other assets.
The vault owner can redeem the collateral by reimbursing the vault with the issued stablecoins (with interest) and is tasked with maintaining the required collateralization.

If the vault becomes undercollateralized, for instance if the price of ETH drops, then an involuntary redemption (liquidation) is performed to deleverage the position.
This deleveraging is performed through buy-backs of stablecoins to close the vault.
Vaults are over-collateralized to help ensure that the position can be closed.
However, should the liquidation proceeds be insufficient, additional mechanisms may kick in to cover the shortfall---either by tapping into a reserve fund or by selling governance tokens as a form of sponsored support (or backstop).
Notably, this occurred in Dai on Black Thursday, when the Maker system found itself in a deleveraging spiral \cite{makerdao2020black,klages2019stability,klages2020while}.

{\it Incentive compatibility.}
Drawing from \cite{werner2021sok}, we consider a cryptoeconomic protocol to be incentive compatible if ``agents are incentivized to execute the game as intended by the protocol designer.''
\cite{klages2020stablecoins} reduces this to a key question of \textit{incentive security}: Is equilibium participation in the stablecoin sustainable?
This requires that incentives among all participants (stablecoin holders, vault owners, governance agents) lead to a mutually profitable equilibrium of participating in the stablecoin.
As non-custodial stablecoins contain self-governing aspects outside of most rule of law, participant incentives are also influenced by the possibility of profitable governance attacks.

{\it This paper.}
We study governance incentive problems in non-custodial stablecoins similar to Maker. We formalize a game theoretic model (an adaptation of that in \cite{klages2020stablecoins}) of governance incentives in Section~\ref{sec:model}.
We derive closed form solutions to stakeholders' problems in Section~\ref{sec:analysis} using financial engineering methods, culminating in conditions for a unique equilibrium in Theorem~\ref{thm:unique-equil}.
We then modify the model to include governance attacks in Section~\ref{sec:gov-attacks} and derive conditions for equilibria without attacks.

\section{Model}\label{sec:model}
We build upon the framework presented in \cite{klages2020stablecoins}, which seeks to describe incentives between governance token holders (GOV), vaults/risk absorbers and stablecoin holders.
We define the model parameters in Table \ref{table_components}.
\begin{table}\centering
\caption{Model components}\label{table_components}
\begin{tabular}{|l|l|}
\hline
Variable &  Definition\\
\hline
        $N$        & Dollar value of vault collateral (COL position)          \\
		$e^{R}$        & Return on COL         \\
		$F$        & Total stablecoin issuance (debt face value)          \\
		$b$        & Return rate on the outside opportunity          \\
		$\beta$        & Collateral factor           \\
		$\delta$        & Interest rate paid by vault to issue STBL          \\
		$u$        & Vault's utility from an outside COL opportunity           \\
		$B$        & STBL market price          \\
\hline
\end{tabular}
\end{table}

We first introduce the basic framework with no attack vectors. The setup considers an interaction between governors and vaults, who both seek to maximize expected profits. The governance choice problem is simply to maximize expected fee revenue. The vault choice problem is to maximize expected revenue from maintaining a long position in COL, while pursuing a new (leveraged) opportunity, and paying an interest fee to governance. 
Randomness is introduced in the model by assuming that the return on COL $e^R$ follows a log-normal distribution:

$$R\sim N(0,\sigma^2),$$ where the standard deviation $\sigma$ represents the COL volatility. We consider continuous compounding returns. The time horizon is set to $1$. In this case
$F(e^\delta - 1)$ represents the total interest paid by vault for the stablecoin issuance, while  $FB(e^b - 1)$ represents the net revenue from investing the proceeds of the stablecoin issuance in the outside opportunity.

Further, vaults are subject to three constraints: Eq.~(\ref{collateral_constraint}), the collateral constraint, which restricts maximum stablecoin issuance to a fraction of posted collateral; Eq.~(\ref{stablecoin_price}), the stablecoin price, which is pegged at one whenever collateral does not fall short; and Eq.~(\ref{outside_utility constraint}), the participation constraint, which simply states that participation must yield higher utility compared to abandoning the system.
Formally, the governance choice problem is written as:
\begin{equation}\label{eq:discrete-gov-system}
	\begin{aligned}
		\max_\delta\quad& \mathbb{E}[(e^\delta - 1)\cdot F]\\
		\text{s.t.}\quad& \text{Vault's choice of $F$.}
	\end{aligned}
\end{equation}
The vault choice problem can be written as:
\begin{align}
	\max_{F\geq0} \hspace{0.2cm} & \mathbb{E}\left[\underbrace{Ne^{R}}_{\text{COL long position}}+\underbrace{F\big(B(e^b - 1)-(e^\delta - 1)\big)}_{\text{Net revenue from leveraged position}}\right] \nonumber \\
	\text{s.t.} \hspace{0.2cm} & \hspace{0.6cm} \text{GOV's choice of $\delta$}\nonumber\\
	&\hspace{0.6cm} F  \leq\beta N \label{collateral_constraint}\\ 
	&\hspace{0.6cm} B  =\mathbb{E}\left[\min\left(1,\frac{Ne^{R}}{F}-(e^\delta - 1) \right)\right]\label{stablecoin_price}\\
	&\hspace{0.6cm} u  \leq\mathbb{E}\left[Ne^{R}+F\big(B(e^b - 1)-(e^\delta - 1)\big)\right].\label{outside_utility constraint}
\end{align}

We consider a Stackelberg equilibrium in which first the governance chooses an interest rate and then the vault chooses the stablecoin issuance. Note that the reverse order would yield a trivial solution in which the vault does not participate.  The governance as a second player would simply set the interest rate $\delta\xrightarrow{}\infty$. In anticipation, vault would set $F=0$ as a consequence of the vaults' participation constraint.
In contrast, the problem in which the governance moves first is non-trivial. Indeed, using financial engineering methods we will give closed form solutions to the objective functions of the two agents. This will allow us to analyze the convexity of their payoffs.
 Under reasonable conditions on parameters, the vaults' participation constraint imposes an upper bound on the interest rate but the optimal interest rate does not saturate this constraint.

\paragraph{ Expected collateral shortfall.}
Our approach follows classical ideas for the valuation of corporate liabilities, present since the seminal works of Black and Scholes \cite{blackscholes79} and Merton \cite{merton1970dynamic}, \cite{merton1974pricing}. ``Since almost all corporate liabilities can be viewed as combinations of options",  i.e., their payoffs can be replicated using an options portfolio, their valuations can be characterized using Black and Scholes formulas, see e.g., \cite{shreve2004stochastic}.

In analogy to the corporate debt holders, the stablecoin holders have an asset essentially equal to $1$ (the face value) minus the following quantity that captures the collateral shortfall:

\begin{equation}
	\begin{aligned}
		P(F,\delta)&=\mathbb{E}\left[Fe^\delta-Ne^R\right]_+=Fe^\delta\Phi(-d_2)-N\Phi(-d_1)\\
		d_1&=\frac{\log(\frac{N}{Fe^\delta})+\frac{\sigma^2}{2}}{\sigma},\qquad d_2=d_1-\sigma.
	\end{aligned}
	\label{putoption}
\end{equation}
Hence, the representative stablecoin holder mimicks the role of the debt holder in classical capital structure models.
The quantity $\Phi(-d_1)$  represents the probability  that there is a collateral shortfall (which is analogous to a corporate default in the classical corporate debt valuation theory): $Fe^\delta > Ne^R$. $\Phi(-d_2)$ is also the probability that there is a shortfall, but adjusted for the severity of this shortfall.\footnote{Note that for the purposes of debt valuation the no-arbitrage theory of option pricing is not relevant. Only the Black and Scholes formulas are needed, i.e. the closed form solution for the expectation in \eqref{putoption} when the random return is log-normal. Moreover, while the valuation of corporate debt can be achieved in a dynamic model, the same formulas govern our one period case where the end of the period can be seen as the bond maturity.}

\paragraph{Vaults' Perspective}
In a similar manner, vaults take into account the expected collateral shortfall in their objective through the stablecoin price constraint,
\begin{equation}
	\begin{aligned}
		\max_F\quad& Ne^\frac{\sigma^2}{2}+F(e^b-e^\delta)-P(F,\delta)(e^b-1)\\
		\text{s.t.}\quad& F\leq \beta\cdot N\\
		&u\leq Ne^\frac{\sigma^2}{2}+F(e^b-e^\delta)-P(\delta,F)(e^b-1)\\
		&\text{GOV's choice of $\delta$.}
	\end{aligned}
\end{equation}

\paragraph{GOV's Perspective}
Governance simply maximizes fee revenue
\begin{equation}
	\begin{aligned}
		\max_\delta\quad& F(e^\delta-1)\\
		\text{s.t.}\quad& \text{Vault's choice of $F$.}
	\end{aligned}
\end{equation}

We later consider an altered form of the model in Section~\ref{sec:gov-attacks} that incorporates a governance attack vector into our analysis.

\section{Stackelberg Equilibrium Analysis}\label{sec:analysis}


Governors  know that the vaults will only choose to participate if the
outside utility of an alternative COL usage is less than (or equal to) the benefit from issuing
stablecoins. We first consider the optimum stablecoin issuance without the outside option constraint, and only with the leverage constraint.
By evaluating vault objective sensitivities (see Appendix~\ref{Derivative_analysis}), we can obtain  vaults' objective maximizer if we include  the leverage constraint (which imposes a cap on the amount of stablecoins issued) but ignore the participation constraint. All proofs are provided in Appendix~\ref{appendix-proofs}.
\begin{proposition}\label{vault_max_unconstrained}
	Vaults' unconstrained objective is maximized at
	\begin{align}
		\varphi(\delta)&= N\cdot\exp\left[\sigma\cdot\Phi^{-1}\left(\frac{e^{b-\delta}-1}{e^b-1}\right)-\delta-\frac{\sigma^2}{2}\right]\label{eq:F-bar},
	\end{align}
	which implicitly requires that $\delta\in[0,b]$.
	Moreover, if we include the leverage constraint, vaults' objective is maximized at
	 \begin{equation}
		F^\ast(\delta)=\min(\varphi(\delta),\beta N).\label{eq:f-star}
	\end{equation}
\end{proposition}
By accounting for vaults' optimal issuance, GOV's objective transforms into
	\begin{equation}
		G= F^\ast(\delta)\left(e^\delta-1\right)\label{eq:g-with-f-star_}.
	\end{equation}

\subsection{\texorpdfstring{$F^\ast(\delta)$}{TEXT} w/o Participation Constraint}
We first derive results disregarding vaults' participation constraint. By comparing the unconstrained optimum stablecoin issuance $\varphi(\delta)$ to $\beta N$ we pin down a lower bound $\delta_{\beta}$ for the interest rate, arising from the leverage constraint, such that $\forall\delta\in( \delta_{\beta},b]$, $\varphi(\delta)<\beta N$. This result is  due to vaults' preference for a larger stablecoin issuance when the interest rate is low, while being unable to exceed the leverage constraint set by governance. Along with the monotonicity of GOV's objective function, we obtain following proposition.

\begin{proposition}[Governance choice]\label{governance_choice_unconstrained}
There exists a $\delta_{\beta}\in[0,b]$ such that $\varphi( \delta_{\beta})=\beta N$. GOV's optimal interest rate, $\delta^{\ast}$, satisfies $\delta^\ast\geq\delta_{\beta}$.
\end{proposition}

Hence, in equilibrium governance will either exhaust the leverage constraint with the highest compatible interest rate or set the interest rate higher than $\delta_\beta$ implying excess overcollateralization. This is not in itself surprising, yet leads us to the following technical lemma.

\begin{lemma}[Concavity threshold for $\delta$]\label{Delta_threshold} There exists a $\delta_\text{th}$ such that for $\delta > \delta_\text{th}$, GOV's objective is concave.
	
\end{lemma}

The following Assumption \ref{assumption-1} ensures that the volatility of the collateral's rate of return is not too large. With this, we further have that GOV's objective is locally increasing at $\delta = \delta_\text{th}$. 
This assumption is currently verified e.g. for ETH.
\begin{assumption}\label{assumption-1}
		$\sigma<2\phi(0)$ (\hyperlink{pf:assumption-1}{See Appendix} for how this condition is derived).
\end{assumption}
By recalling that $\frac{dG}{d\delta}$ is non-increasing with $\delta$ for  $\delta > \delta_{th}$ (from Lemma \ref{Delta_threshold}), we then obtain the following proposition.

\begin{proposition}[Governance unconstrained optimal choice]\label{Delta_upper}
	Under Assumption \ref{assumption-1} there exists a $\delta^{\ast}$, at which level GOV's objective is maximized. Consequently, if there exists $\delta_\text{th}< \delta_{\beta}<\delta^{\ast}\leq b$, then GOV would take $\delta^{\ast}$.
	
\end{proposition}

Then $\delta=\delta^{\ast}$ achieves the unconstrained maximum for the GOV objective. For $\delta_\text{th}< \delta_{\beta}$ to hold, we need the following assumption.
\begin{assumption}\label{beta-assumption}
        $\beta<\frac{e^b+1}{2}\cdot\exp(-b-\frac{\sigma^2}{2})$.
\end{assumption}
\noindent This is because $\varphi( \delta_{\beta})=\beta N$ and $\frac{d\varphi}{d\delta}<0$, when we ask for $\delta_\text{th}< \delta_{\beta}$, we therefore need to have $\varphi(\delta_\text{th})>\varphi( \delta_{\beta})=\beta N$. Intuitively,  governance will achieve a lower payoff at the interest rate pinned down by the leverage constraint, $\delta_{\beta}$, relative to a lower interest rate, $\delta_\text{th}$, which would, in turn, allow vaults to issue a larger amount of stablecoins resulting in larger interest revenue for governance ceteris paribus.
Note that the RHS at Assumption~\ref{beta-assumption} obtains its maximum value of $\exp(-\frac{\sigma^2}{2})<1$ when $b=0$, implying overcollateralization.

\subsection{\texorpdfstring{$F^\ast(\delta)$}{TEXT} w/ Participation Constraint}
We now give conditions on the parameters for which the optimal interest rate set by governors satisfies the vaults' participation constraint given outside utility on COL usage. First, we make the following additional assumption. 
\begin{assumption}\label{u-assumption}
		    $u\leq Ne^\frac{\sigma^2}{2}+N\Phi(-d_1(\delta^{\ast}))(e^b-1)$.
\end{assumption}
Here we ensure that the vault is able to achieve a payoff equal to or above their utility from an outside COL opportunity. This assumption ensures that GOV is aware that their optimum interest rate must be sufficiently attractive in order for vaults to participate, i.e.,  governance must take into account the vaults' idiosyncratic tradeoffs. Armed with this assumption, we characterize a unique equilibrium in the following theorem.
\begin{theorem}\label{thm:unique-equil}
	If hyper-parameters are selected such that Assumption \ref{assumption-1}, \ref{beta-assumption}, and \ref{u-assumption} all are satisfied, and there exist $ \delta_{\beta}$ and $\delta^{\ast}$ that satisfy \eqref{eq:delta-lower_} and \eqref{eq:delta-upper} respectively, then there is a unique equilibrium with
	$\delta=\delta^{\ast}$ and $F =\varphi(\delta^{\ast})$.
\end{theorem}
\section{Governance attack vector}\label{sec:gov-attacks}
We now introduce a governance attack vector, as per \cite{gudgeon2020decentralized}. 
A rational adversary only engages in an attack if its profits exceed costs.
They could exploit the governance system to change the contract code and access a sufficiently large GOV token stake to approve the update.
For instance, the adversarial change could transfer all COL to the adversary's address.
More nuanced versions can also extract collateral indirectly by manipulating price feeds as in \cite{klagesmundt2019vuln}.
This may not require 50\% of GOV tokens as governance participation is commonly low.
Neither does it require a single wealthy adversary, since many attackers can collude via a crowdfunding strategy (e.g., \cite{daian2018}), or a single attacker could borrow the required tokens via a flash loan (as in \cite{zoltu2019,gudgeon2020decentralized}). Note that timelocks make it harder to pursue flash loan governance attacks.

Formally, we consider an adversarial agent with a $\zeta$ fraction of GOV tokens, who is able to steal a $\gamma$ fraction of collateral in the system. Typically, we might have $\gamma=1$, although not always. A rational attack will take place when profits exceed costs, i.e., when
$
    \zeta F\frac{e^\delta-1}{1-r}+\alpha<\gamma \mathbb{E}[Ne^R]=\gamma Ne^\frac{\sigma^2}{2},
$
where $\alpha$ is an outside cost to attack, and $\zeta F\frac{e^\delta-1}{1-r}$ is the opportunity cost of an attack, i.e. the profits resulting from a non-attack decision, represented as a geometric series of future fee revenue with discount factor $r$. 
In idealized DeFi, we might have $\alpha=0$ or very close to 0 (through pseudonymity), while, in traditional finance, $\alpha$ is assumed to be so high such that an attack would always be unprofitable, e.g. due to legal repercussions. 


We could extend the vault choice problem to include the amount of collateral, $N$, locked in the stablecoin system as a share of total endowed collateral available to the vault, $\bar{N}$. Only locked in collateral is subject to seizure during a governance attack. 
We assume for simplicity that $\bar{N} = N$ and we leave the decision on how much collateral to lock in as an open problem.
If there is no attack, i.e., $\alpha+\zeta F\frac{e^\delta-1}{1-r} \geq\gamma Ne^\frac{\sigma^2}{2}$,  the governance choice problem writes as before in \eqref{eq:discrete-gov-system}
and if there is an attack, i.e., $\alpha+\zeta F\frac{e^\delta-1}{1-r} < \gamma Ne^\frac{\sigma^2}{2}$, then the governance's payoff (ex-adversary) is equal to zero.

In the Stackelberg equilibrium with vaults as a second player, the vault choice problem is only relevant conditional on the governance attack being unsuccessful, in which case it writes as before.
If the attack is successful, then there is no participation from vaults and the stablecoin system is abandoned. 
We are thus interested in the non-attack scenario with participation from the vault, since only then is there mutually profitable continued participation for both parties and we have incentive security.

The optimal interest rate set by governance that ensures a non-attack decision (and so incentive security) then satisfies the following condition:
\begin{equation}
	\alpha+\zeta \underbrace{F(\delta^\ast)\frac{e^{\delta^\ast}-1}{1-r}}_{G^\ast} \geq \gamma Ne^\frac{\sigma^2}{2}
	\hspace{0.5cm} \text{equivalently, } G^\ast \geq\frac{\gamma N^\frac{\sigma^2}{2}-\alpha}{\zeta},
	\label{eq:incentive_security}
\end{equation}
where $G^*$ is the optimal governance  objective value (and $\delta^\ast$ is the unique optimizing interest rate) under the assumptions of Theorem \ref{thm:unique-equil}.

$$
\frac{\gamma N e^{\frac{\sigma^2}{2}}-\alpha}{\zeta}
$$

%

Since $\delta^\ast$ represents a Stackelberg equilibrium with vault participation, condition \eqref{eq:incentive_security} is both necessary and sufficient  for the existence of an interest  rate that satisfies both the non-attack condition \textit{and} the participation constraint.
Note that an interest rate  that satisfies  condition \eqref{eq:incentive_security} may not be feasible in general. The practical implication of the condition is that participants in the system can use it to verify the incentives of decentralized governors and whether given conditions lead to an equilibrium with incentive security or whether governors may have perverse incentives. The condition applies given \emph{rational} behaviour, since agents know ex-ante if an attack will take place based on parameter values.

\section{Conclusion}
We have characterized the unique equilibrium arising in non-custodial stablecoins with decentralized governance.
The payoff structure is based on closed form  valuations of the positions of  two stakeholders in the capital structure that underlies the stablecoin.
We obtain the equilibrium interest rate and level of participation in settings without governance attacks (Theorem~\ref{thm:unique-equil}) and with a possible governance attack (Eq.~\ref{eq:incentive_security}).
Using these closed form solutions, protocol designers can more easily account for the effects their design choices will have on economic equilibrium and incentive security in the system. Our results allow us to quantify how loose the participation constraint can be in order  to allow governors to earn a sufficiently high profit in the stablecoin system such that it offsets the proceeds from attacking the system. The implication is that GOV tokens should be  expensive enough (e.g., from fundamental value of `honest' cash flows) so that it is unprofitable for outsiders to buy them with the sole purpose of attacking the system.

By comparing the precise value of the GOV tokens to the return of the collateral at stake, adjusted for the attack cost, we can evaluate the security and sustainability of decentralized governance systems.
As the adjusted attack cost increases with $\alpha$, one possible mitigation to strengthen these governance systems is the traditional one: increase $\alpha$ to deter attack through centralized means. One  way to do this to make governors resemble legal fiduciaries with known identities, which often goes against the idealized tenets of DeFi. Another possibility, recently proposed in \cite{lee2021gov} as ``optimistic approval'', alters the problem in a different way by incorporating a veto mechanism invokable by other parties in the system (e.g., vaults and stablecoin holders) in the case of malicious governance proposals. This would introduce a new term in our model that lowers the success probability of an attack based on the probability that the veto mechanism is invoked. If governors anticipate that the veto mechanism will be invoked, then their expectations of attack profit plummet, expanding the mutual participation region. 

\paragraph*{Acknowlegements.}
The authors thank the Center for Blockchains and Electronic Markets at University of Copenhagen for support.

\bibliographystyle{splncs04}

\appendix

\section{Derivative Analysis}\label{Derivative_analysis}
\subsection{Sensitivity of the Expected Collateral Shortfall}
Note the following relationship,
\begin{align}
	Fe^\delta\phi(d_2)=N\phi(d_1)
\end{align}
With this, we have the following derivatives,
\begin{align}
	\frac{\partial P}{\partial F}&=e^\delta\Phi(-d_2)+Fe^\delta\cdot\phi(-d_2)\cdot\left(-\frac{d d_2}{dF}\right)-N\cdot\phi(-d_1)\cdot\left(-\frac{d d_1}{dF}\right)\nonumber\\
	&=e^\delta\Phi(-d_2)\\
	\frac{\partial P}{\partial\delta}&=Fe^\delta\Phi(-d_2)+Fe^\delta\cdot\phi(-d_2)\cdot\left(-\frac{d d_2}{d\delta}\right)-N\cdot\phi(-d_1)\cdot\left(-\frac{d d_1}{d\delta}\right)\nonumber\\
	&=Fe^\delta\Phi(-d_2)
\end{align}
\subsection{Vault Objective Sensitivities}
Denote
\begin{equation}
	V:=\quad Ne^\frac{\sigma^2}{2}+F(e^b-e^\delta)-P(\delta,F)(e^b-1)\label{eq:V}
\end{equation}
Note the following derivatives,
\begin{align}
	\frac{\partial V}{\partial F}&=\left(e^b-e^\delta\right)-\left(e^b-1\right)e^\delta\Phi(-d_2)\label{eq:dv-df}\\
	\frac{\partial V}{\partial\delta}&=-Fe^\delta-(e^b-1)\frac{\partial P}{\partial\delta}\nonumber\\
	&=-Fe^\delta-(e^b-1)Fe^\delta\Phi(-d_2)\nonumber\\
	&=-Fe^\delta(1+(e^b-1)\delta\Phi(-d_2))<0\quad\text{always}
\end{align}

\subsection{GOV Objective Sensitivities}
Denote
\begin{equation}
	G:=\quad F\left(e^\delta-1\right)\label{eq:G}
\end{equation}
Note the following derivatives,
\begin{align}
	\frac{\partial G}{\partial\delta}&=Fe^\delta>0\quad\text{always}\\
	\frac{\partial G}{\partial F}&=e^\delta-1
\end{align}

\section{Proofs}\label{appendix-proofs} 

\textbf{Proposition \ref{vault_max_unconstrained}}\hypertarget{pf:vault-max-unconstrained}{}
\begin{proof}
	Since $V$ is concave in $F$, we set  \eqref{eq:dv-df} equal to zero to obtain a maximum for $V$ :
	\begin{align}
		\Phi(-d_2)&=\frac{e^b-e^\delta}{e^\delta\left(e^b-1\right)}\nonumber\\
		\frac{\log\left(\frac{F}{N}\right)+\delta+\frac{\sigma^2}{2}}{\sigma}&=\Phi^{-1}\left(\frac{e^{b-\delta}-1}{e^b-1}\right)\nonumber\\
		F^\ast=\varphi(\delta)&= N\cdot\exp\left[\sigma\cdot\Phi^{-1}\left(\frac{e^{b-\delta}-1}{e^b-1}\right)-\delta-\frac{\sigma^2}{2}\right], 
	\end{align}
	with \eqref{eq:F-bar} implicitly requiring that $\delta\in[0,b]$.\\
	Together with the leverage constraint, this implies
	\begin{equation}
		F^\ast(\delta)=\min(\varphi(\delta),\beta N) 
	\end{equation}
	since the leverage constraint imposes a cap on amount of stablecoins issued.
	Substitute \eqref{eq:f-star} into \eqref{eq:G} and obtain
	\begin{equation}
		G= F^\ast(\delta)\left(e^\delta-1\right)\label{eq:g-with-f-star},
	\end{equation}
	thus transforming GOV's optimization into finding the optimum for \eqref{eq:g-with-f-star}.
\end{proof}

\noindent\textbf{Proposition \ref{governance_choice_unconstrained}}\hypertarget{pf:governance-choice-unconstrained}{}
\begin{proof}

We begin by establishing a lower bound for $\delta$:
	There exists a $\delta_{\beta}\in[0,b]$ such that $\varphi( \delta_{\beta})=\beta N$, i.e.
	\begin{equation}
		\frac{F^{\ast}}{N}=\exp\left[\sigma\cdot\Phi^{-1}\left(\frac{e^{b- \delta_{\beta}}-1}{e^b-1}\right)- \delta_{\beta}-\frac{\sigma^2}{2}\right]=\beta\label{eq:delta-lower_}.
	\end{equation}
	The quantity $\delta_{\beta}$ is the  interest rate for which vaults' leverage constraint is hit, i.e. for $\delta < \delta_{\beta}$ the optimal stablecoin issuance is given by $\beta N$.

	Indeed, by comparing $\varphi(\delta)$ to $\beta N$ we obtain
	\begin{align}
		\frac{d\varphi}{d\delta}&=-\varphi(\delta)\cdot\left[1+\sigma\cdot\frac{1}{\phi\left(\Phi^{-1}\left(\frac{e^{b-\delta}-1}{e^b-1}\right)\right)}\cdot\frac{e^{b-\delta}}{e^b-1}\right]<0\quad\text{always}\label{eq:dvarphi-ddelta}
	\end{align}
	Thus $\exists \delta_{\beta}\in[0,b]$ such that $\varphi( \delta_{\beta})=\beta N$, i.e.
	\begin{equation}
		\frac{F^{\ast}}{N}=\exp\left[\sigma\cdot\Phi^{-1}\left(\frac{e^{b- \delta_{\beta}}-1}{e^b-1}\right)- \delta_{\beta}-\frac{\sigma^2}{2}\right]=\beta\label{eq:delta-lower},
	\end{equation}
	which effectively is setting a lower bound to $\delta$, 
	such that $\forall\delta\in( \delta_{\beta},b]$, $\varphi(\delta)<\beta N$.

We can no conclude the proof of Proposition \ref{governance_choice_unconstrained}.
\textbf{Suppose $\varphi(\delta)\geq\beta N$}, i.e. $\delta\in[0, \delta_{\beta}]$
	\begin{align*}
		G&=\beta N\cdot\left(e^\delta-1\right)\\
		\frac{dG}{d\delta}&=\beta Ne^\delta>0\quad\text{always}.
	\end{align*}
	Thus, GOV will choose $\delta^\ast= \delta_{\beta}$. 
\end{proof}

\noindent\textbf{Lemma \ref{Delta_threshold}}\hypertarget{pf:delta-threshold}{}
\begin{proof}
	\textbf{Suppose $\varphi(\delta)<\beta N$}, i.e. $\delta\in( \delta_{\beta},b]$
	\begin{align*}
		G&=\varphi(\delta)\cdot\left(e^\delta-1\right)\\
		\frac{dG}{d\delta}&=\varphi(\delta)\cdot e^\delta + \frac{\partial\varphi}{\partial\delta}\cdot(e^\delta-1)\\
		&=\varphi(\delta)\cdot e^\delta -\varphi(\delta)\cdot\left[1+\sigma\cdot\frac{1}{\phi\left(\Phi^{-1}\left(\frac{e^{b-\delta}-1}{e^b-1}\right)\right)}\cdot\frac{e^{b-\delta}}{e^b-1}\right]\cdot(e^\delta-1)\\
		&=\varphi(\delta)\left[1-\sigma\cdot\frac{1}{\phi\left(\Phi^{-1}\left(\frac{e^{b-\delta}-1}{e^b-1}\right)\right)}\cdot\frac{e^b-e^{b-\delta}}{e^b-1}\right]
	\end{align*}
	from which the lemma follows. Consider threshold value $\delta_\text{th}$ such that 
	\begin{align*}
		\frac{e^{b-\delta_\text{th}}-1}{e^b-1}&=0.5\quad\Rightarrow\quad \delta_\text{th}=b-\log\left(\frac{e^b+1}{2}\right).
	\end{align*}
	When $\delta>\delta_\text{th}$, we have as $\delta$ increases
	\begin{itemize}
		\item $\Phi^{-1}\left(\frac{e^{b-\delta}-1}{e^b-1}\right)$ decreases from $0$ to $-\infty$
		\item $\phi\left(\Phi^{-1}\left(\frac{e^{b-\delta}-1}{e^b-1}\right)\right)$ hence decreases from $\phi(0)$ to 0
		\item $\frac{1}{\phi\left(\Phi^{-1}\left(\frac{e^{b-\delta}-1}{e^b-1}\right)\right)}$ increases from $\frac{1}{\phi(0)}$ to $\infty$
		\item $\frac{e^b-e^{b-\delta}}{e^b-1}$ increases from 0.5 to $\frac{e^b}{e^b-1}$
		\item Overall, $\sigma\cdot\frac{1}{\phi\left(\Phi^{-1}\left(\frac{e^{b-\delta}-1}{e^b-1}\right)\right)}\cdot\frac{e^b-e^{b-\delta}}{e^b-1}$ is increasing.
	\end{itemize}
\end{proof}

\noindent\textbf{Assumption \ref{assumption-1}}\hypertarget{pf:assumption-1}{}
\begin{proof}
	\begin{align}
		&1-\sigma\cdot\frac{1}{\phi\left(\Phi^{-1}\left(\frac{e^{b-\delta_\text{th}}-1}{e^b-1}\right)\right)}\cdot\frac{e^b-e^{b-\delta_\text{th}}}{e^b-1}>0\nonumber\\
		\Leftrightarrow\quad&1-\frac{\sigma}{2\phi(0)}>0\nonumber\\ \Leftrightarrow\quad&\sigma<2\phi(0)
	\end{align}
\end{proof}

\noindent\textbf{Proposition \ref{Delta_upper}}\hypertarget{pf:delta-upper}{}
\begin{proof}
	Under Assumption \ref{assumption-1}, we have that $G$ is locally increasing at $\delta =\delta_{th}$ and we have that  $\frac{dG}{d\delta}$ is non-increasing with $\delta$ for  $\delta > \delta_{th}$. 
	
	Therefore, when setting
	$\frac{dG}{d\delta}=0$, i.e. $1-\sigma\cdot\frac{1}{\phi\left(\Phi^{-1}\left(\frac{e^{b-\delta}-1}{e^b-1}\right)\right)}\cdot\frac{e^b-e^{b-\delta}}{e^b-1}=0$, we implicitly obtain a $\delta^{\ast}$, at which level $G$ is maximized.
\end{proof}

\noindent\textbf{Theorem \ref{thm:unique-equil}}\hypertarget{pf:unique-equil}{}
\begin{proof}
	
	At $\delta^{\ast}$, we have
	
	\begin{equation}
		\sigma\cdot\frac{1}{\phi\left(\Phi^{-1}\left(\frac{e^{b-\delta^{\ast}}-1}{e^b-1}\right)\right)}\cdot\frac{e^b-e^{b-\delta^{\ast}}}{e^b-1}=1\label{eq:delta-upper}
	\end{equation}

	Substitute \eqref{eq:delta-upper} into \eqref{eq:V},
	\begin{equation}
		\begin{aligned}
		V(\delta^{\ast})&=Ne^\frac{\sigma^2}{2}+\varphi(\delta^{\ast})(e^b-e^{\delta^{\ast}})\nonumber\\
		&-P(\delta^{\ast},\varphi(\delta^{\ast}))(e^b-1)\label{eq:v-delta-upper}\\
		\end{aligned}
	\end{equation}
	\begin{equation}
		\begin{aligned}
		P(\delta^{\ast},\varphi(\delta^{\ast}))&=\varphi(\delta^{\ast})e^{\delta^{\ast}}\Phi(-d_2)-N\Phi(-d_1)\label{eq:p-delta-upper}\\
		\end{aligned}	
	\end{equation}
	\begin{equation}
		\begin{aligned}
		d_1(\delta^{\ast})&=\frac{1}{\delta}\cdot\left(\log(\frac{N}{\varphi(\delta^{\ast})e^\delta})+\frac{\sigma^2}{2}\right)\nonumber\\
		&=-\Phi^{-1}\left(\frac{e^{b-\delta^{\ast}}-1}{e^b-1}\right)+\sigma\label{eq:d1-delta-upper}\\
		\end{aligned}
	\end{equation}
	\begin{equation}
		\begin{aligned}
		d_2(\delta^{\ast})&=d_1-\sigma=-\Phi^{-1}\left(\frac{e^{b-\delta^{\ast}}-1}{e^b-1}\right)\label{eq:d2-delta-upper}
		\end{aligned}
	\end{equation}
	
	We obtain
	\begin{align*}
		P(\delta^{\ast},\varphi(\delta^{\ast}))&=\varphi(\delta^{\ast})\frac{e^{b}-e^{\delta^{\ast}}}{e^b-1}-N\Phi(-d_1)\label{eq:p-delta-upper-sub}
	\end{align*}
	
	and
	\begin{align*}
		V(\delta^{\ast})
		&=Ne^\frac{\sigma^2}{2}+N\Phi(-d_1)(e^b-1)
	\end{align*}
	such that we must assume 
	\begin{equation}
		u\leq Ne^\frac{\sigma^2}{2}+N\Phi(-d_1(\delta^{\ast}))(e^b-1),
	\end{equation}
	in order for vault participation to hold.
\end{proof}


\end{document}